\documentclass[10pt]{article}

\usepackage{amsmath}
\usepackage{amssymb}

\usepackage{graphicx}
\usepackage{lscape}

\usepackage{cite}

\usepackage{color} 

\usepackage{setspace} 
\usepackage[labelfont=bf,labelsep=period,justification=raggedright]{caption}

\makeatletter
\renewcommand{\@biblabel}[1]{\quad#1.}
\makeatother

\date{}

\begin{document}

% Title must be 150 characters or less
\begin{flushleft}
{\Large
\textbf{Power laws in citation distributions: Evidence from Scopus}
}
% Insert Author names, affiliations and corresponding author email.
\\
\bigskip
Michal Brzezinski
\\
\bf Faculty of Economic Sciences, University of Warsaw, Warsaw, Poland 
\\
E-mail: mbrzezinski@wne.uw.edu.pl
\end{flushleft}

% Please keep the abstract between 250 and 300 words
\section*{Abstract}

Modeling distributions of citations to scientific papers is crucial for understanding how science develops. However, there is a considerable empirical controversy on which statistical model fits the citation distributions best. This paper is concerned with rigorous empirical detection of power-law behaviour in the distribution of citations received by the most highly cited scientific papers. We have used a large, novel data set on citations to scientific papers published between 1998 and 2002 drawn from Scopus. The power-law model is compared with a number of alternative models using a likelihood ratio test. We have found that the power-law hypothesis is rejected for around half of the Scopus fields of science. For these fields of science, the Yule, power-law with exponential cut-off and log-normal distributions seem to fit the data better than the pure power-law model. On the other hand, when the power-law hypothesis is not rejected, it is usually empirically indistinguishable from most of the alternative models. The pure power-law model seems to be the best model only for the most highly cited papers in ``Physics and Astronomy''. Overall, our results seem to support theories implying that the most highly cited scientific papers follow the Yule, power-law with exponential cut-off or log-normal distribution. Our findings suggest also that power laws in citation distributions, when present, account only for a very small fraction of the published papers (less than 1\% for most of science fields) and that the power-law scaling parameter\ (exponent) is substantially higher (from around 3.2 to around 4.7) than found in the older literature.

% Please keep the Author Summary between 150 and 200 words
% Use first person. PLoS ONE authors please skip this step. 
% Author Summary not valid for PLoS ONE submissions.   
%\section*{Author Summary}

\section*{Introduction}
\label{intro}
It is often argued in scientometrics, social physics and other sciences that distributions of some scientific "items" (e.g., articles, citations) produced by some scientific sources (e.g., authors, journals) have heavy tails that can be modelled using a power-law model. These distributions are then said to conform to the Lotka's law \cite{lotka1926frequency}. Examples of such distributions include author productivity, occurrence of words, citations received by papers, nodes of social networks, number of authors per paper, scattering of scientific literature in journals, and many others \cite{egghe2005power}. In fact, power-law models are widely used in many sciences as physics, biology, earth and planetary sciences, economics, finance, computer science, and others \cite{newman2005power,clauset2009power}. Models equivalent to Lotka's law are known as Pareto's law in economics \cite{gabaix2009power} and as Zipf's law in linguistics \cite{baayen2001word}. Appropriate measuring and providing scientific explanations for power laws plays an important role in understanding the behaviour of various natural and social phenomena.

This paper is concerned with empirical detection of power-law behaviour in the distribution of citations received by scientific papers. The power-law distribution of citations for the highly cited papers was first suggested by Price \cite{price1965}, who also proposed a ``cumulative advantage'' mechanism that could generate the power-law distribution \cite{price1976general}. More recently, a growing literature has developed that aims at measuring power laws in the right tails of citation distributions. In particular, Redner \cite{redner1998popular,redner2005citation} found that the right tails of citation distributions for articles published in Physical Review over a century and of articles published in 1981 in journals covered by Thomson Scientific's Web of Science (WoS) follow power laws. The latter data set was also modelled with power-law techniques by Clauset et al. \cite{clauset2009power} and Peterson et al. \cite{peterson2010nonuniversal}. The latter study also used data from 2007 list of the living highest h-index 
chemists and from Physical Review D between 1975 and 1994.  Van Raan \cite{van2006statistical}  observed that the top of the distribution of around 18,000 papers published between 1991 and 1998 in the field of chemistry in Netherlands follows a power law distribution. Power-law models were also fitted  to data from high energy physics \cite{lehmann2003citation}, data for most cited physicists \cite{laherrere1998stretched}, data for all papers published in journals of the American Physical Society from 1983 to 2008\cite{eom2011characterizing}, and to data for all physics papers published between 1980 and 1989 \cite{golosovsky2012runaway}. 

Recently, Albarr\'{a}n and Ruiz-Castillo \cite{albarran2011references} tested for the  power-law behavior using a large WoS dataset of 3.9 million articles published between 1998 and 2002 categorized in 22 WoS research fields. The same dataset was also used to search for the power laws in the right tail of citation distributions categorized in 219 WoS scientific sub-fields \cite{albarran2011skewness,albarran2011skewness_2}. These studies offer the largest existing body of evidence on the power-law behaviour of citation distributions. Three major conclusions appear from them. First, the power-law behavior is not universal. The existence of power law cannot be rejected in the WoS data for 17 out of 22 and for 140 out of 219 sub-fields studied in 
\cite{albarran2011references} and in \cite{albarran2011skewness,albarran2011skewness_2}, respectively. Secondly, in opposition to previous studies, these papers found that the scaling parameter (exponent)\ of the power-law distribution is above 3.5 in most of the cases, while the older literature suggested that the parameter value is between 2 and 3 \cite{albarran2011skewness_2}. Third, power laws in citation distributions are rather small --– on average they cover just about 2\% of the most highly cited articles in a given WoS field of science and account for about 13.5\% of all citations in the field.

The main aim of this paper is to use a statistically rigorous approach to answer the empirical question of whether the power-law model describes best the observed distribution of highly cited papers. We use the statistical toolbox for detecting power-law behaviour introduced by Clauset et al. \cite{clauset2009power}. There are two major contributions of the present paper. First, we use a very large, previously unused data set on the citation distributions of the most highly cited papers in several fields of science. This data set comes from Scopus, a bibliographic database introduced in 2004 by Elsevier, and contains 2.2 million articles
published between 1998 and 2002 and categorized in 27 Scopus major subject areas of science. Most of the previous studies used rather small data sets, which were not suitable for rigorous statistical detecting of the power-law behaviour. In contrast, our sample is even bigger with respect to the most highly cited papers than the large sample used in the recent contributions based on WoS data \cite{albarran2011references,albarran2011skewness,albarran2011skewness_2}.
This results from the fact that Scopus indexes about 70\% more sources compared to the WoS\cite{lopez2008coverage,aghaei2013comparison} and therefore gives a more comprehensive coverage of citation distributions.\footnote{From the perspective of measuring power laws in citation distributions, the most important part of the distribution is the right tail. It seems that the database used in this paper has a better coverage of the right tail of citation distributions. The most highly cited paper in our database has received 5187 citations (see Table \ref{tab:table2}), while the corresponding number for the database based on WoS is 4461 \cite{li2013impact}. Our database is further described in ``Materials and methods'' section.}

The second major contribution of the paper is to provide a rigorous statistical comparison of the power-law model and a number of alternative models with respect to the problem which theoretical distribution fits better empirical data on citations. This problem of model selection has been previously studied in some contributions to the literature. It has been argued that  models like stretched exponential \cite{laherrere1998stretched}, Yule \cite{price1976general}, log-normal \cite{redner2005citation,stringer2008effectiveness,radicchi2008universality}, Tsallis \cite{tsallis2000citations,anastasiadis2010tsallis,wallace2009modeling} or shifted power law \cite{eom2011characterizing} fit citation distributions equally well or better than the pure power-law model. However, previous papers have either focused on a single alternative distribution or used only visual methods to choose between the competing models. The present paper fills the gap  by providing a 
systematic and statistically rigorous comparison of the power-law distribution with such alternative models as the log-normal, exponential, stretched exponential (Weibull), Tsallis, Yule and power-law with exponential cut-off. The comparison between models was performed using a likelihood ratio test \cite{vuong1989likelihood,clauset2009power}.   

\section*{Materials and Methods}

\subsection*{Fitting power-law model to citation data}
\label{sec:fitting}

We follow Clauset et al. \cite{clauset2009power} in choosing methods for fitting power laws to citation distributions. These authors carefully show that, in general, the appropriate methods depend on whether the data are continuous or discrete. In our case, the latter is true as citations are non-negative integers. Let $x$ be the number of citations received by an article in a given field of science. The probability density function (pdf) of the discrete power-law model is defined as
\begin{equation}
p(x)= \frac{x^{-\alpha}}{\zeta(\alpha,x_0)},
\label{eq:pl_def}
\end{equation}
where $\zeta(\alpha,x_0)$ is the generalized or Hurwitz zeta function. The $\alpha$ is a shape parameter of the power-law distribution, known as the power-law exponent or scaling parameter. The power-law behaviour is usually found only for values greater than some minimum, denoted by $x_0$. In case of citation distributions, the power-law behaviour has been found on average only in the top 2\% of all articles published in a field of science \cite{albarran2011skewness, albarran2011skewness_2}. 

The lower bound on the power-law behaviour, $x_0$, should be therefore estimated if we want to measure precisely in which part of a citation distribution the model applies. Moreover, we need an estimate of $x_0$ if we want to obtain an unbiased estimate of the power-law exponent, $\alpha$. 

We estimate $\alpha$ using the maximum likelihood (ML) estimation. The log-likelihood function corresponding to (\ref{eq:pl_def}) is
\begin{equation}
L(\alpha) = -n\ln\zeta(\alpha,x_0) - \alpha\sum_{i=1}^{n}\ln x_i
\label{eq:ll}
\end{equation}

The ML estimate for $\alpha$ is found by numerical maximization of (2).\footnote{Clauset at al. \cite{clauset2009power} provide also an approximate method of estimating $\alpha$ for the discrete power-law model by assuming that continuous power-law distributed reals are rounded to the nearest integers. However, it this paper we use an exact approach based on maximizing (\ref{eq:ll}).}

Following Clauset et al. \cite{clauset2009power}, we use the following procedure to estimate the lower bound on the power-law behaviour, $x_{0}$. For each $x\geqslant x_{min}$, we calculate the ML\ estimate of the power-law exponent, $\hat\alpha$, and then we compute the well-known Kolmogorov-Smirnov (KS) statistic for the data and the fitted model. The KS statistic is defined as
\begin{equation}
\mathrm{KS} = \max\limits_{x\geqslant x_0} \left\vert S(x) - P(x; \hat\alpha)\right\vert,
\end{equation}
where $S(x)$ is the cumulative distribution function (cdf) for the observations with value at least $x_0$, and $P(x,\hat\alpha)$ is the cdf for the fitted power-law model to observations for which $x\geqslant x_0$. The estimate $\hat{x}_0$ is then chosen as a value of $x_0$ for which the KS statistic is the smallest. The standard errors for both estimated parameters, $\hat\alpha$ and $\hat{x}_0$, are computed with standard bootstrap methods with 1,000 replications. 

\subsection*{Goodness-of-fit and model selection tests}
\label{sec:tests}

The next step in measuring power laws involves testing goodness of fit. A positive result of such a test allows to conclude that a power-law model is consistent with data. Following Clauset et al. \cite{clauset2009power} again, we use a test based on a semi-parametric bootstrap approach.\footnote{If our data were drawn from a given model, then we could use the KS statistic in testing goodness of fit, because the distribution of the KS statistic is known in such a case. However, when the underlying model is not known or when its parameters are estimated from the data, which is our case, the distribution of the KS statistic must be obtained by simulation.}
 The procedure starts with fitting a power-law model to data and calculating a KS statistic for this fit, $k$. Next, a large number of synthetic data sets is generated that follow the originally fitted power-law model above the estimated $x_{0}$ and have the same non-power-law distribution as the original data set below $\hat x_{0}$. Then, a power-law model is fitted to each of the generated data sets using the same methods as for the original data set, and the KS statistics are calculated.
The fraction of data sets for which their own KS statistic is larger than $k$ is  the \textit{p}-value of the test. It represents a probability that the KS statistics computed for data drawn from the power-law model fitted to the original data is
at least as large as $k$.
The power-law hypothesis is rejected if the $p$-value is smaller than some chosen threshold. Following Clauset et al. \cite{clauset2009power}, we rule out the power-law model if the estimated $p$-value for this test is smaller than 0.1. In the present paper, we use 1,000 generated  data sets. 

If  the goodness-of-fit test rejects the power-law hypothesis, we may conclude that the power law has not been found. However, if a data set is fitted well by a power law, the question remains if there is an alternative distribution, which is an equally good or better fit to this data set. We need, therefore, to fit some rival distributions and evaluate which distribution gives a better fit. To this aim, we use the likelihood ratio test, which tests if the compared models are equally close to the true model against the alternative that one is closer. The test computes the logarithm of the ratio of the likelihoods of the data under two competing distributions, LR, which is negative or positive depending on which model fits data better. Specifically, let us consider two distributions with pdfs denoted by $p_1(x)$ and $p_2(x)$. The LR is defined as:
\begin{equation}
\mathrm{LR} = \sum_{i=1}^{n}[\ln p_1(x_i)-\ln p_2(x_i)].
\end{equation}
A positive value of the LR suggests that model $p_1(x)$ fits the data better.
However, the sign of the LR can be used to determine which model should be favored only if the LR is significantly different from zero. Vuong \cite{vuong1989likelihood} showed that in the case of non-nested models the normalized
log-likelihood ratio $\mathrm{NLR}=n^{-1/2}\mathrm{LR}/\sigma$, where $\sigma$ is the estimated standard deviation of LR, has a limit standard normal distribution.\footnote{In case of nested models, $2\mathrm{LR}$ has a limit a chi-squared distribution \cite{vuong1989likelihood}.} This result can be used to compute a \textit{p}-value for the test discriminating between the competing models. If the \textit{p}-value is small (for example, smaller than 0.1), then the sign of the LR can probably be trusted as an indicator of which model is preferred. However, if the \textit{p}-value is large, then the test is unable to choose between the compared distributions.

We have followed Clauset et al. \cite{clauset2009power} in choosing the following alternative discrete distributions: exponential, stretched exponential (Weibull), log-normal, Yule and the power law with exponential cut-off.\footnote{The power-law with exponential cut-off behaves like the pure power-law model for smaller values of $x$, $x\geqslant x_0$,while for larger values of $x$ it behaves like an exponential distribution. The pure power-law model is nested within the power-law with exponential cut-off, and for this reason the latter always provides a fit at least as good as the former.} Most of these models have been considered in previous literature on modeling citation distribution. As another alternative, we also use the Tsallis distribution, which has been also proposed as a model for citation distributions \cite{wallace2009modeling,anastasiadis2010tsallis}.
The definitions of our alternative distributions are given in Table \ref{tab:table1}.
\begin{center}
[Please insert Table \ref{tab:table1} about here]
\end{center}

\subsection*{Data}
\label{data}

We use citation data from Scopus, a bibliographic database introduced in 2004 by Elsevier. Scopus is a major competitor to the most-widely used data source in the literature on modeling citation distributions -- Web of Science (WoS) from Thomson Reuters. Scopus covers 29 million records with references going back to 1996 and 21 million pre-1996 records going back as far as 1823. An important limitation of the database is that it does not cover cited references for pre-1996 articles. Scopus contains 21,000 peer-reviewed journals from more than 5,000 international publishers. It covers about 70\% more sources compared to the
WoS \cite{lopez2008coverage}, but a large part of the additional sources are low-impact journals. A recent literature review has found that the quite extensive literature that compares WoS and Scopus from the perspective of citation analysis offers mixed results \cite{aghaei2013comparison}. However, most of the studies suggest that, at least for the period from 1996 on, the number of citations in both databases is either roughly similar or higher in Scopus than in WoS. Therefore, is seems that Scopus constitutes a useful alternative to WoS from the perspective of modeling citation distributions.

Journals in Scopus are classified under four main subject areas: life sciences (4,200 journals), health sciences (6,500 journals), physical sciences (7,100 journals) and social sciences including arts and humanities (7,000 journals). The four main subject areas are further divided into 27 major subject areas and more than 300 minor subject areas. Journals may be classified under more than one subject area.

The analysis in this paper was performed on the level of 27 Scopus major subject areas of science.\footnote{See Table \ref{tab:table2} for a list of the analyzed Scopus areas of science.} From the various document types contained in Scopus, we have selected only articles. For the purpose of comparability with the recent WoS-based studies \cite{albarran2011references,albarran2011skewness}, only the articles published between 1998 and 2002 were considered. Following previous literature, we have chosen a common 5-year citation window for all articles published in 1998-2002.\footnote{For example, for articles published in 1998 we have analyzed citations received during 1998-2002, while for articles published in 2002, those received during 2002-2006.} See  Albarr\'{a}n and Ruiz-Castillo \cite{albarran2011references} for a justification of choosing the 5-year citation window common for all fields of science. 

In order to measure the power-law behaviour of citations, we need data on the right tails of citation distributions.
To this end, we have used the Scopus Citation Tracker to collect citations for $\min(100,000; x)$ of the highest cited articles, where $x$ is the actual number of articles published in a given field of science during 1998-2002. This analysis was performed separately for each of the 27 science fields categorized by Scopus. 

Descriptive statistics for our data sets are presented in Table \ref{tab:table2}.
\begin{center}[Please insert Table \ref{tab:table2} about here]\end{center}
In some cases, there was less than 100,000 articles published in a field of science during 1998-2002 and we were able to obtain complete or almost complete distributions of citations (see columns 2-4 of Table \ref{tab:table2}).\footnote{For all fields of science analyzed, there were some articles with missing information on citations. These articles were removed from our samples. However, this has usually affected only about 0.1\% of our samples.} In other cases, we have obtained only a part of the relevant distribution encompassing the right tail and some part of the middle of the distribution. The smallest portions of citation distributions were obtained for Medicine (8.4\% of total papers), Biochemistry, Genetics and Molecular Biology (15.7\%) and Physics and Astronomy (18.4\%). However, using the WoS data for 22 science categories,  Albarr\'{a}n and Ruiz-Castillo \cite{albarran2011references} found that power laws account usually only for less than 2\% of the highest-cited articles. Therefore, it seems that the coverage of the right tails of citation distributions in our samples is satisfactory for our purposes.

% Results and Discussion can be combined.
\section*{Results and Discussion}

\label{results}

\begin{figure}[!ht]
\begin{center}
\includegraphics[scale=0.8]{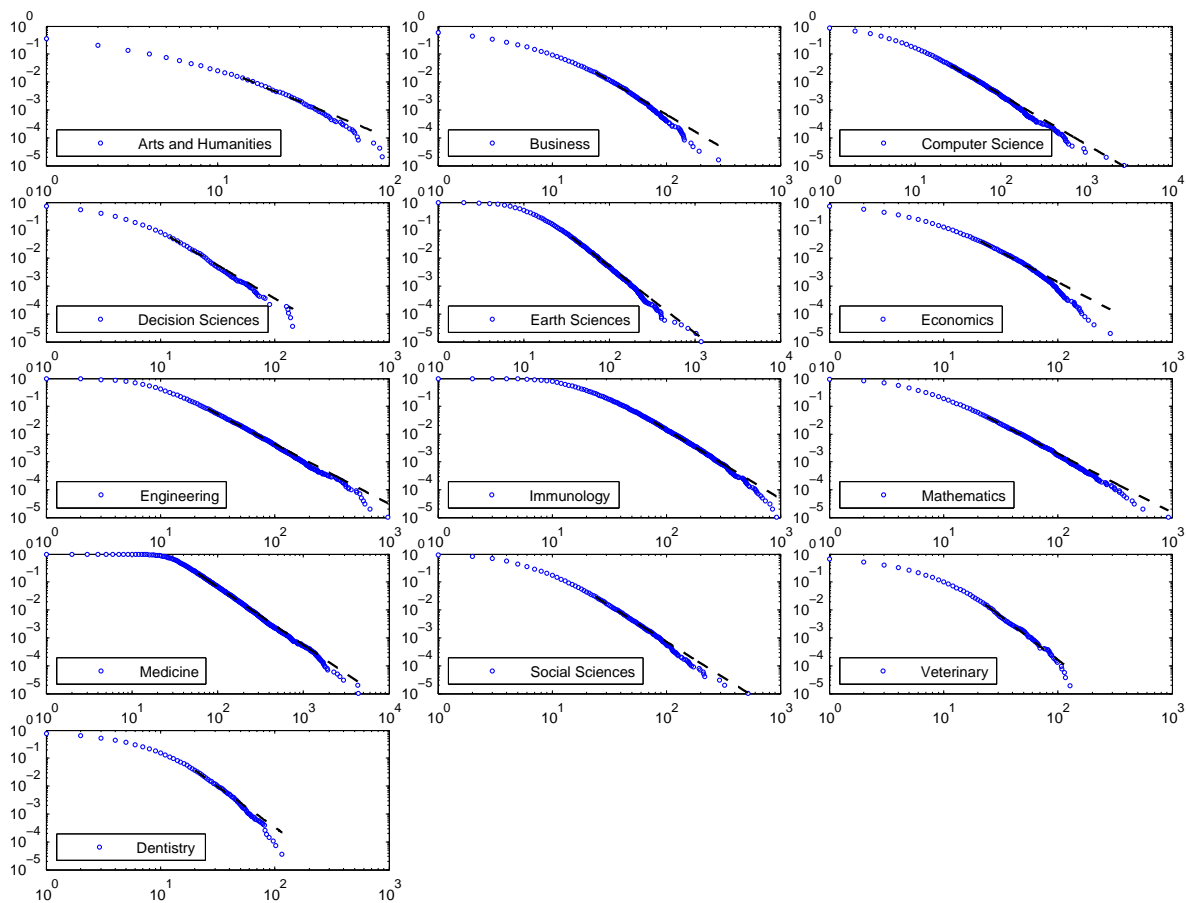}
\end{center}
\caption{\bf The complementary cumulative distribution functions (blue circles) and best power-law fits (dashed black line) for citation distributions that did not pass the goodness-of-fit test, Scopus, 1998--2002, 5-year citation window.} 
\label{figure1}
\end{figure}

Table \ref{tab:table3} presents results of fitting the discrete power-law model to our data sets consisting of citations to scientific articles published over 1998-2002 (with a common 5-year citation window), separately for each of the 27 Scopus major subject areas of science. The last row gives also results for all subject areas combined (``All sciences'). Beside estimates of the power-law exponent ($\hat{\alpha}$) and the lower bound on the power-law behaviour ($\hat{x}_0$), the table gives also the estimated number and the percentage of power-law distributed papers, as well as the \textit{p}-value for our goodness-of-fit test. \begin{center}[Please insert Table \ref{tab:table3} about here]\end{center}
Results with respect to the goodness-of-fit suggest that the power-law hypothesis cannot be rejected for the following 14 Scopus science fields: ``Agricultural and Biological Sciences'', ``Biochemistry, Genetics and Molecular Biology'', ``Chemical Engineering'', ``Chemistry'', ``Energy'', ``Environmental Science'', ``Materials Science'', ``Neuroscience'', ``Nursing'', ``Pharmacology, Toxicology and Pharmaceutics'', ``Physics and Astronomy'', ``Psychology'', ``Health Professions'', and ``Multidisciplinary''. The remaining 13 Scopus fields of science for which the power-law model is rejected include humanities and social sciences (``Arts and Humanities'', ``Business, Management and Accounting'', ``Economics, Econometrics and Finance'', ``Social Sciences''), but also formal sciences (``Computer Science'', ``Decision Sciences'', ``Mathematics''), life sciences (``Immunology and Microbiology'', ``Medicine'', ``Veterinary'', ``Dentistry''),  as well as ``Earth and Planetary Sciences'' and ``Engineering''. The best power-law fits for these fields of science are shown on Figure  \ref{figure1}.  
\begin{center}[Please insert Figure \ref{figure1} about here]\end{center}
For most of the distributions shown on Figure \ref{figure1}, it can be clearly seen that their right tails decay faster than the pure power-law model indicates. This suggest that the largest observations for these distributions should be rather modeled with a distribution having a lighter tail than the pure power-law model like the log-normal or power-law with exponential cut-off models. 

The \textit{p}-value for our goodness-of-fit test in case of ``All Sciences''  is 0.076, which is below our acceptance threshold of 0.1. However, this \textit{p}-value is non-negligible and significantly higher than \textit{p}-values for most of the 13 Scopus fields of science for which we reject the power-law hypothesis. For this reason, we conclude that the evidence is not conclusive in this case. Our result for ``All Sciences'' is, however, in a stark contrast with that of Albarr\'{a}n and Ruiz-Castillo \cite{albarran2011references}, who using the WoS data found that the fit for a corresponding data set was very good (with a \textit{p}-value of 0.85).\footnote{In Albarr\'{a}n and Ruiz-Castillo \cite{albarran2011references}, the power-law hypothesis is found plausible for 17 out of 22 WoS fields of science. It is rejected for ``Pharmacology and Toxicology", ``Physics'', ``Agricultural Sciences'', ``Engineering'', and ``Social Sciences, General''. These results are not directly comparable with those of the present paper as Scopus and WoS use different classification systems to categorize journals.} 

The estimates of the power-law exponent for the 14 Scopus science fields for which the power law seems to be a plausible hypothesis range from 3.24 to 4.69. This is in a good agreement with Albarr\'{a}n and Ruiz-Castillo \cite{albarran2011references} and confirms their assessment that the true value of this parameter is substantially higher than found in the earlier literature \cite{redner1998popular,lehmann2003citation,tsallis2000citations}, which offered estimates ranging from around 2.3 to around 3. We also confirm the observation of Albarr\'{a}n and Ruiz-Castillo \cite{albarran2011references} that power laws in citation distributions are rather small -- they account usually for less than 1\% of total articles published in a field of science. The only two fields in our study with slightly ``bigger'' power laws are ``Chemistry'' (2\%) and ``Multidisciplinary'' (2.8\%).

The comparison between the power-law hypothesis and alternatives using the Vuong's test is presented in Table \ref{tab:table4}. It can be observed that the exponential model can be ruled out in most of the cases. We discuss other results first for the 13 Scopus fields of science that did not pass our goodness-of-fit test. For all of these fields, except for ``Veterinary'', the Yule and power-law with exponential cut-off models fit the data better than the pure power-law model in a statistically significant way. The log-normal model is better than the pure power-law model in 10 of the discussed fields; the same holds for the Weibull distribution in case of 5 fields. However, these results do not imply that the distributions, which give a better fit to the non-power-law distributed data than the pure power-law model are plausible hypotheses for these data sets. This issue should be further studied using appropriate goodness-of-fit tests. 
\begin{center}[Please insert Table \ref{tab:table4} about here]\end{center}
We now turn to results for the remaining Scopus fields of science that were not rejected by our goodness-of-fit test. The power-law hypothesis seems to be the best model only for ``Physics and Astronomy''. In this case, the test statistics is always non-negative implying that the power-law model fits the data as good as or better than each of the alternatives. For the remaining 13 fields of science, the log-normal, Yule and power-law with exponential cut-off models have always higher log-likelihoods suggesting that these models may fit the data better than the pure power-law distribution. However, only in a few cases the differences between models are statistically significant. For ``Chemistry'' and ``Multidisciplinary'' both the Yule and power-law with exponential cut-off models are favoured over the pure power-law model. The power-law with exponential cut-off is also favoured in case of ``Health Professions''. In other cases, the \textit{p}-values for the likelihood ratio test are large, which implies that there is no conclusive evidence that would allow to distinguish between the pure power-law, log-normal, Yule and power-law with exponential cut-off distributions. Comparing the power-law distribution with the Weibull and Tsallis distributions, we observe that the sign of the test statistics is positive in roughly half of the cases, but the \textit{p}-values are always large and neither model can be ruled out. Our likelihood ratio tests suggest therefore that when the power-law is a plausible hypothesis according to our goodness-of-fit test it is often indistinguishable from
alternative models.  

Overall, our results show that the evidence in favour of the power-law behaviour of the right-tails of citation distributions is rather weak. For roughly half of the Scopus fields of science studied, the power-law hypothesis is rejected. Other distributions, especially the Yule, power-law with exponential cut-off and log-normal seem to fit the data from these fields of science better than the pure power-law model. On the other hand, when the power-law hypothesis is not rejected, it is usually empirically indistinguishable from all alternatives with the exception of the exponential distribution. The pure power-law model seems to be favoured over alternative  models only for the most highly cited papers in ``Physics and Astronomy''. Our results suggest that theories implying that the most highly cited scientific papers follow the Yule, power-law with exponential cut-off or log-normal distribution may have slightly more support in data than theories predicting the pure power-law behaviour. 
 
\section*{Conclusions}

We have used a large, novel data set on citations to scientific papers published between 1998 and 2002 drawn from Scopus to test empirically for the power-law behaviour of the right-tails of citation distributions. We have found that the power-law hypothesis is rejected for around half of the Scopus fields of science. For the remaining fields of science, the power-law distribution is a plausible model, but the differences between the power law and alternative models are usually statistically insignificant.
The paper also confirmed recent findings of Albarr´an and Ruiz-Castillo [17] that power laws in citation distributions, when they are a plausible, account only for a very small fraction of the published papers (less than 1\% for most of science fields) and that the power-law exponent is substantially higher than found in the older literature. %). 

% Do NOT remove this, even if you are not including acknowledgments
\section*{Acknowledgments}
I would like to acknowledge gratefully the use of Matlab and R software accompanying the papers by Clauset et al. \cite{clauset2009power} and Shalizi \cite{shalizi2007maximum}. Any remaining errors are my responsibility. 

%\section*{References}
% The bibtex filename
\bibliography{pl_citations}

\begin{thebibliography}{10}
\providecommand{\url}[1]{\texttt{#1}}
\providecommand{\urlprefix}{URL }
\expandafter\ifx\csname urlstyle\endcsname\relax
  \providecommand{\doi}[1]{doi:\discretionary{}{}{}#1}\else
  \providecommand{\doi}{doi:\discretionary{}{}{}\begingroup
  \urlstyle{rm}\Url}\fi
\providecommand{\bibAnnoteFile}[1]{%
  \IfFileExists{#1}{\begin{quotation}\noindent\textsc{Key:} #1\\
  \textsc{Annotation:}\ \input{#1}\end{quotation}}{}}
\providecommand{\bibAnnote}[2]{%
  \begin{quotation}\noindent\textsc{Key:} #1\\
  \textsc{Annotation:}\ #2\end{quotation}}
\providecommand{\eprint}[2][]{\url{#2}}

\bibitem{lotka1926frequency}
Lotka A (1926) The frequency distribution of scientific productivity.
\newblock Journal of Washington Academy Sciences 16: 317--323.
\bibAnnoteFile{lotka1926frequency}

\bibitem{egghe2005power}
Egghe L (2005) Power laws in the information production process: Lotkaian
  informetrics.
\newblock Oxford: Elsevier.
\bibAnnoteFile{egghe2005power}

\bibitem{newman2005power}
Newman ME (2005) {Power laws, Pareto distributions and Zipf's law}.
\newblock Contemporary Physics 46: 323--351.
\bibAnnoteFile{newman2005power}

\bibitem{clauset2009power}
Clauset A, Shalizi CR, Newman ME (2009) Power-law distributions in empirical
  data.
\newblock SIAM review 51: 661--703.
\bibAnnoteFile{clauset2009power}

\bibitem{gabaix2009power}
Gabaix X (2009) Power laws in economics and finance.
\newblock Annual Review of Economics 1: 255--294.
\bibAnnoteFile{gabaix2009power}

\bibitem{baayen2001word}
Baayen RH (2001) Word frequency distributions.
\newblock Dordrecht: Kluwer.
\bibAnnoteFile{baayen2001word}

\bibitem{price1965}
de~Solla~Price D (1965) Networks of scientific papers.
\newblock Science 149: 510--515.
\bibAnnoteFile{price1965}

\bibitem{price1976general}
de~Solla~Price D (1976) A general theory of bibliometric and other cumulative
  advantage processes.
\newblock Journal of the American Society for Information Science 27: 292--306.
\bibAnnoteFile{price1976general}

\bibitem{redner1998popular}
Redner S (1998) {How popular is your paper? An empirical study of the citation
  distribution}.
\newblock The European Physical Journal B 4: 131--134.
\bibAnnoteFile{redner1998popular}

\bibitem{redner2005citation}
Redner S (2005) {Citation statistics from 110 years of Physical Review}.
\newblock Physics Today 58: 49--54.
\bibAnnoteFile{redner2005citation}

\bibitem{peterson2010nonuniversal}
Peterson GJ, Press{\'e} S, Dill KA (2010) Nonuniversal power law scaling in the
  probability distribution of scientific citations.
\newblock Proceedings of the National Academy of Sciences 107: 16023--16027.
\bibAnnoteFile{peterson2010nonuniversal}

\bibitem{van2006statistical}
Van~Raan AF (2006) Statistical properties of bibliometric indicators: Research
  group indicator distributions and correlations.
\newblock Journal of the American Society for Information Science and
  Technology 57: 408--430.
\bibAnnoteFile{van2006statistical}

\bibitem{lehmann2003citation}
Lehmann S, Lautrup B, Jackson A (2003) Citation networks in high energy
  physics.
\newblock Physical Review E 68: 026113.
\bibAnnoteFile{lehmann2003citation}

\bibitem{laherrere1998stretched}
Laherr{\`e}re J, Sornette D (1998) Stretched exponential distributions in
  nature and economy:"fat tails" with characteristic scales.
\newblock The European Physical Journal B 2: 525--539.
\bibAnnoteFile{laherrere1998stretched}

\bibitem{eom2011characterizing}
Eom YH, Fortunato S (2011) Characterizing and modeling citation dynamics.
\newblock PloS One 6: e24926.
\bibAnnoteFile{eom2011characterizing}

\bibitem{golosovsky2012runaway}
Golosovsky M, Solomon S (2012) Runaway events dominate the heavy tail of
  citation distributions.
\newblock The European Physical Journal Special Topics 205: 303--311.
\bibAnnoteFile{golosovsky2012runaway}

\bibitem{albarran2011references}
Albarr{\'a}n P, Ruiz-Castillo J (2011) References made and citations received
  by scientific articles.
\newblock Journal of the American Society for Information Science and
  Technology 62: 40--49.
\bibAnnoteFile{albarran2011references}

\bibitem{albarran2011skewness}
Albarr{\'a}n P, Crespo JA, Ortu{\~n}o I, Ruiz-Castillo J (2011) The skewness of
  science in 219 sub-fields and a number of aggregates.
\newblock Scientometrics 88: 385--397.
\bibAnnoteFile{albarran2011skewness}

\bibitem{albarran2011skewness_2}
Albarr{\'a}n P, Crespo JA, Ortu{\~n}o I, Ruiz-Castillo J (2011) The skewness of
  science in 219 sub-fields and a number of aggregates.
\newblock Working paper 11-09, Universidad Carlos III.
\bibAnnoteFile{albarran2011skewness_2}

\bibitem{lopez2008coverage}
L{\'o}pez-Illescas C, de~Moya-Aneg{\'o}n F, Moed HF (2008) {Coverage and
  citation impact of oncological journals in the Web of Science and Scopus}.
\newblock Journal of Informetrics 2: 304--316.
\bibAnnoteFile{lopez2008coverage}

\bibitem{aghaei2013comparison}
Aghaei~Chadegani A, Salehi H, Md~Yunus M, Farhadi H, Fooladi M, et~al. (2013)
  {A comparison between two main academic literature collections: Web of
  Science and Scopus databases}.
\newblock Asian Social Science 9: 18--26.
\bibAnnoteFile{aghaei2013comparison}

\bibitem{li2013impact}
Li Y, Ruiz-Castillo J (2013) The impact of extreme observations in citation
  distributions.
\newblock Technical report, Universidad Carlos III, Departamento de
  Econom{\'\i}a.
\bibAnnoteFile{li2013impact}

\bibitem{stringer2008effectiveness}
Stringer MJ, Sales-Pardo M, Amaral LAN (2008) Effectiveness of journal ranking
  schemes as a tool for locating information.
\newblock PLoS One 3: e1683.
\bibAnnoteFile{stringer2008effectiveness}

\bibitem{radicchi2008universality}
Radicchi F, Fortunato S, Castellano C (2008) Universality of citation
  distributions: Toward an objective measure of scientific impact.
\newblock Proceedings of the National Academy of Sciences 105: 17268--17272.
\bibAnnoteFile{radicchi2008universality}

\bibitem{tsallis2000citations}
Tsallis C, de~Albuquerque MP (2000) Are citations of scientific papers a case
  of nonextensivity?
\newblock The European Physical Journal B 13: 777--780.
\bibAnnoteFile{tsallis2000citations}

\bibitem{anastasiadis2010tsallis}
Anastasiadis AD, de~Albuquerque MP, de~Albuquerque MP, Mussi DB (2010) {Tsallis
  q-exponential describes the distribution of scientific citations - a new
  characterization of the impact}.
\newblock Scientometrics 83: 205--218.
\bibAnnoteFile{anastasiadis2010tsallis}

\bibitem{wallace2009modeling}
Wallace ML, Larivi{\`e}re V, Gingras Y (2009) Modeling a century of citation
  distributions.
\newblock Journal of Informetrics 3: 296--303.
\bibAnnoteFile{wallace2009modeling}

\bibitem{vuong1989likelihood}
Vuong QH (1989) Likelihood ratio tests for model selection and non-nested
  hypotheses.
\newblock Econometrica 57: 307--333.
\bibAnnoteFile{vuong1989likelihood}

\bibitem{shalizi2007maximum}
Shalizi CR (2007) {Maximum likelihood estimation for q-exponential (Tsallis)
  distributions}.
\newblock Technical report, arXiv preprint math/0701854.
\bibAnnoteFile{shalizi2007maximum}

\end{thebibliography}

\clearpage

\section*{Figure Legends}
%\begin{figure}[!ht]
%\begin{center}
%%\includegraphics[width=4in]{figure_name.2.eps}
%\end{center}
%\caption{
%{\bf Bold the first sentence.}  Rest of figure 2  caption.  Caption 
%should be left justified, as specified by the options to the caption 
%package.
%}
%\label{Figure_label}
%\end{figure}

\clearpage

\section*{Tables}
%\begin{table}[!ht]
%\caption{
%\bf{Table title}}
%\begin{tabular}{|c|c|c|}
%table information
%\end{tabular}
%\begin{flushleft}Table caption
%\end{flushleft}
%\label{tab:label}
% \end{table}

\begin{table}[!ht]
\caption{\bf Definitions of alternative discrete distributions.} 
\begin{tabular}{lc} \hline
Distribution name & Probability distribution function \\ \hline
Exponential &   $(1-\mathrm{e}^{-\lambda})\mathrm{e}^{\lambda x_0}\mathrm{e}^{-\lambda x}$ \\
Stretched exponential (Weibull) & $\frac{1}{\sum_{x_0}^{\infty}(q^{x^\beta} - q^{(x+1)^\beta})} q^{x^\beta} - q^{(x+1)^\beta}$ \\

Log-normal  &   $\sqrt{\frac{2}{\pi\sigma^2}}\left[\mathrm{erfc}(\frac{\mathrm{ln}x_{0}-\mu}{\sqrt{2}\sigma})\right]^{-1} \frac{1}{x}\exp\left[ -\frac{(\mathrm{ln}x-\mu)^2}{2\sigma^2} \right]$ \\ Tsallis & $\frac{1}{\sum_{x_0}^{\infty}(1+x/\sigma)^{-\theta-1}}(1+x/\sigma)^{- \theta -1} $ \\ 
Yule        &  $(\alpha -1)\frac{\Gamma(x_0+\alpha-1)}{\Gamma(x_0)} \frac{\Gamma(x)}{\Gamma(x+\alpha)}$  \\
Power law with exponential cut-off & $(\sum_{x_0}^{\infty} x^{-\alpha} \mathrm{e}^{-\lambda x})^{-1} x^{-\alpha}\mathrm{e}^{-\lambda x}$ \\
\hline
\end{tabular}
\begin{flushleft}
Note: The distributions have been normalized to ensure that the total probability in the domain $[x_0,+\infty]$ is 1. Discrete log-normal distribution is approximated by rounding the continuous log-normally distributed reals to the nearest integers. For Tsallis distribution, we use a parametrization considered by Shalizi \cite{shalizi2007maximum}. 
\end{flushleft}
\label{tab:table1}
\end{table}

\begin{landscape}
\begin{table}
\centering
\footnotesize
\caption{\bf Descriptive statistics for citation distributions, Scopus, 1998--2002, 5-year citation window}
\begin{tabular}{lcccccc}
\hline 
 Scopus subject area of science  & Total number & No. of papers & \% of all papers & Mean no. & Std. Dev.  & Max. no. \\ 
   & of  papers & in the sample & in the sample & of citations & of citations & of citations \\ \hline  
Agricultural and Biological Sciences & 372575 & 99804 & 26.8 & 15.17 & 14.36 & 628 \\  
Arts and Humanities & 47191 & 47074 & 99.8 & 1.256 & 3.357 & 91 \\  
Biochemistry, Genetics and Molecular Biology & 636421 & 99819 & 15.7 & 49.09 & 46.29 & 3118 \\  
Business, Management and Accounting & 61211 & 61156 & 99.9 & 3.452 & 7.273 & 287 \\  
Chemical Engineering & 158673 & 98989 & 62.4 & 7.232 & 9.236 & 344 \\  
Chemistry & 416660 & 99398 & 23.9 & 21.07 & 21.17 & 1065 \\  
Computer Science & 134179 & 99933 & 74.5 & 6.44 & 18.13 & 2737 \\  
Decision Sciences  & 27409 & 27393 & 99.9 & 3.467 & 5.496 & 143 \\  
Earth and Planetary Sciences & 228197 & 99788 & 43.7 & 14.1 & 17.03 & 1195 \\  
Economics, Econometrics and Finance & 49645 & 49559 & 99.8 & 4.652 & 8.653 & 287 \\  
Energy & 67076 & 66378 & 99.0 & 2.553 & 5.596 & 334 \\  
Engineering & 439719 & 99765 & 22.7 & 11.77 & 15.83 & 971 \\  
Environmental Science & 186898 & 99847 & 53.4 & 10.72 & 11.27 & 730 \\  
Immunology and Microbiology & 195339 & 99858 & 51.1 & 22.11 & 25.11 & 926 \\  
Materials Science & 331310 & 99591 & 30.1 & 12.48 & 14.49 & 697 \\  
Mathematics & 193740 & 99922 & 51.6 & 6.912 & 11.38 & 929 \\  
Medicine & 1191154 & 99823 & 8.4 & 48.55 & 60.14 & 4365 \\  
Neuroscience & 445181 & 99886 & 22.4 & 18.97 & 20.39 & 771 \\  
Nursing & 51283 & 50464 & 98.4 & 5.274 & 12.07 & 518 \\  
Pharmacology, Toxicology and Pharmaceutics & 179427 & 99757 & 55.6 & 12.19 & 12.28 & 347 \\  
Physics and Astronomy & 541328 & 99817 & 18.4 & 24.75 & 31.64 & 3118 \\  
Psychology & 104449 & 99736 & 95.5 & 7.446 & 11.55 & 377 \\  
Social Sciences & 215410 & 99890 & 46.4 & 6.148 & 8.055 & 519 \\  
Veterinary & 53203 & 53117 & 99.8 & 3.637 & 5.843 & 128 \\  
Dentistry & 27470 & 27437 & 99.9 & 4.943 & 6.736 & 115 \\  
Health Professions & 75491 & 75414 & 99.9 & 7.272 & 11.49 & 348 \\  
Multidisciplinary & 50287 & 50226 & 99.9 & 30.38 & 76.08 & 5187 \\  
All Sciences  & 6480926 & 2203841 & 34.0 & 14.92 & 27.74 & 5187 \\  \hline
\end{tabular}
\label{tab:table2}
\end{table}
\end{landscape}

\begin{landscape}
\begin{table}
\centering
\footnotesize
\caption{\bf Power-law fits to citation distributions, Scopus, 1998--2002, 5-year citation window}
\begin{tabular}{lccccc}
\hline
Scopus subject area of science & $\hat{x}_0$ & $\hat{\alpha}$ &  No. of power-law papers & \% of total papers & $p$--value \\
\hline
Agricultural and Biological Sciences & 92 (15.1) & 4.19(0.25) & 488 & 0.1 & 0.566 \\  
Arts and Humanities & 14 (5.4) & 3.46 (0.47) & 655 & 1.4 & 0.005 \\  
Biochemistry, Genetics and Molecular Biology & 148 (28.0) & 3.72 (0.13) & 2813 & 0.4 & 0.175 \\  
Business, Management and Accounting & 24 (10.1) & 3.4 (0.38) & 1339 & 2.2 & 0.000 \\  
Chemical Engineering & 38 (6.7) & 4.01 (0.19) & 1418 & 0.9 & 0.099 \\  
Chemistry & 41(7.1) & 3.4(0.05) & 8193 & 2.0 & 0.110 \\  
Computer Science & 26 (10.6) & 2.78 (0.11) & 3989 & 3.0 & 0.000 \\  
Decision Sciences  & 12 (4.0) & 3.36 (0.24) & 1596 & 5.8 & 0.000 \\  
Earth and Planetary Sciences & 36 (8.9) & 3.37 (0.09) & 5834 & 2.6 & 0.000 \\  
Economics, Econometrics and Finance & 21 (10.2) & 3.13 (0.36) & 1995 & 4.0 & 0.000 \\  
Energy & 32 (5.4) & 3.91 (0.22) & 356 & 0.5 & 0.825 \\  
Engineering & 26 (9.4) & 3.14 (0.09) & 7986 & 1.8 & 0.000 \\  
Environmental Science & 63 (10.3) & 4.33 (0.22) & 624 & 0.3 & 0.506 \\  
Immunology and Microbiology & 78 (13.6) & 3.48 (0.10) & 2713 & 1.4 & 0.049 \\  
Materials Science & 43 (8.9) & 3.47 (0.11) & 2687 & 0.8 & 0.193 \\  
Mathematics & 24 (4.0) & 3.11 (0.06) & 4152 & 2.1 & 0.012 \\  
Medicine & 59 (16.3) & 3.07 (0.04) & 20163 & 1.7 & 0.000 \\  
Neuroscience & 135 (28.4) & 4.69 (0.41) & 423 & 0.1 & 0.896 \\  
Nursing & 60 (15.7) & 3.68 (0.40) & 439 & 0.9 & 0.256 \\  
Pharmacology, Toxicology and Pharmaceutics & 56 (6.8) & 4.1 (0.12) & 1215 & 0.7 & 0.865 \\  
Physics and Astronomy & 61 (6.5) & 3.35 (0.04) & 5034 & 0.9 & 0.797 \\  
Psychology & 52 (8.8) & 3.9 (0.17) & 1060 & 1.0 & 0.812 \\  
Social Sciences & 24 (6.4) & 3.56 (0.15) & 2963 & 1.4 & 0.007 \\  
Veterinary & 23 (4.0) & 4.09 (0.27) & 858 & 1.6 & 0.017 \\  
Dentistry & 20 (2.4) & 3.89 (0.18) & 1012 & 3.7 & 0.011 \\  
Health Professions & 49 (10.2) & 3.85 (0.24) & 942 & 1.2 & 0.352 \\  
Multidisciplinary & 209 (40.4) & 3.24 (0.14) & 1147 & 2.8 & 0.100 \\  
All Sciences  & 186 (46.3) & 3.45 (0.10) & 6364 & 0.2 & 0.076 \\  \hline
\end{tabular}
\begin{flushleft}
Note: standard errors are given in parentheses.
\end{flushleft}
\label{tab:table3}
\end{table}
\end{landscape}

\begin{landscape}
\begin{table}
\centering

\footnotesize
\caption{\bf Model selection tests for citation distributions, Scopus, 1998--2002, 5-year citation window}

\begin{tabular}{lccccccccccccc}  \hline
Scopus subject area of science & \textit{p}-value & \multicolumn{2}{c}{Exponential} & \multicolumn{2}{c}{Weibull} & \multicolumn{2}{c}{Log-normal} & \multicolumn{2}{c}{Tsallis} & \multicolumn{2}{c}{\newline Yule} & \multicolumn{2}{c}{PL with cut-off}  \\  
 & \textit{} & LR & \textit{p} & LR & \textit{p} & LR & \textit{p} & LR & \textit{p} & LR & \textit{p} & NLR & \textit{p} \\  \hline
Agricultural and Biological Sciences & 0.566 & 20.740 & 0.009 & 0.338 & 0.779 & -0.096 & 0.782 & 0.054 & 0.890 & -0.011 & 0.858 & -0.268 & 0.464  \\  
Arts and Humanities & 0.005 & 6.287 & 0.457 & -6.93 & 0.023 & -6.56 & 0.025 & -4.325 & 0.189 & -1.38 & 0.000 & -7.37 & 0.000  \\  
Biochemistry, Genetics and Molecular Biology & 0.175 & 204.5 & 0.000 & 1.22 & 0.758 & -1.12 & 0.473 & -1.227 & 0.479 & -0.155 & 0.108 & -0.567 & 0.287\\  
Business, Management and Accounting & 0.000 & 34.390 & 0.034 & -9.60 & 0.013 & -9.24 & 0.013 & -7.279 & 0.065 & -1.39 & 0.000 & -9.98 & 0.000  \\  
Chemical Engineering & 0.099 & 69.480 & 0.001 & -0.021 & 0.994 & -0.972 & 0.480 & 0.025 & 0.990 & -0.358 & 0.187 & -0.78 & 0.211  \\  
Chemistry & 0.110 & 736.0 & 0.000 & 7.48 & 0.262 & -2.67 & 0.204 & 1.290 & 0.687 & -0.999 & 0.060 & -3.31 & 0.010  \\  
Computer Science & 0.000 & 609.4 & 0.000 & -7.05 & 0.248 & -8.80 & 0.035 & -6.719 & 0.132 & -2.00 & 0.000 & -5.23 & 0.001  \\  
Decision Sciences  & 0.000 & 77.730 & 0.001 & -6.71 & 0.046 & -6.81 & 0.048 & -.0275 & 0.956 & -2.66 & 0.000 & -5.91 & 0.001  \\  
Earth and Planetary Sciences & 0.000 & 459.7 & 0.000 & -4.69 & 0.451 & -7.52 & 0.045 & -4.928 & 0.264 & -1.95 & 0.000 & -5.69 & 0.001   \\  
Economics, Econometrics and Finance & 0.000 & 45.080 & 0.021 & -21.6 & 0.000 & -20.4 & 0.000 & -17.027 & 0.002 & -2.68 & 0.000 & -22.9 & 0.000  \\  
Energy & 0.825 & 20.630 & 0.065 & 0.357 & 0.789 & -0.072 & 0.838 & 0.347 & 0.690 & -0.023 & 0.884 & -0.119 & 0.625  \\  
Engineering & 0.000 & 825.5 & 0.000 & - & - & -7.98 & 0.032 & -0.763 & 0.877 & -2.71 & 0.000 & -7.52 & 0.000   \\  
Environmental Science & 0.506 & 26.730 & 0.104 & 0.003 & 0.999 & -0.422 & 0.685 & -0.333 & 0.793 & -0.114 & 0.334 & -0.18 & 0.547  \\  
Immunology and Microbiology & 0.049 & 170.3 & 0.000 & -1.85 & 0.539 & -2.48 & 0.176 & -1.111 & 0.496 & -0.268 & 0.076 & -3.98 & 0.005  \\  
Materials Science & 0.193 & 233.4 & 0.000 & 2.02 & 0.610 & -1.02 & 0.460 & -0.034 & 0.987 & -0.412 & 0.178 & -0.850 & 0.192   \\  
Mathematics & 0.012 & 414.8 & 0.000 & -1.54 & 0.784 & -4.97 & 0.083 & -0.264 & 0.943 & -1.56 & 0.007 & -5.19 & 0.001   \\  
Medicine & 0.000 & 2740.0 & 0.000 & - & - & -7.78 & 0.043 & -4.566 & 0.309 & -2.03 & 0.000 & -5.62 & 0.001   \\  
Neuroscience & 0.896 & 11.920 & 0.072 & -0.018 & 0.987 & -0.178 & 0.726 & -0.066 & 0.888 & -0.020 & 0.637 & -0.285 & 0.451   \\  
Nursing & 0.256 & 21.520 & 0.012 & -0.284 & 0.803 & -0.372 & 0.580 & -0.048 & 0.936 & -0.045 & 0.565 & -0.733 & 0.226  \\  
Pharmacology, Toxicology and Pharmaceutics & 0.865 & 47.520 & 0.000 & -0.361 & 0.844 & -0.747 & 0.449 & -0.002 & 0.999 & -0.148 & 0.337 & -1.24 & 0.115 \\  
Physics and Astronomy & 0.797 & 706.2 & 0.000 & 19.5 & 0.006 & 0.048 & 0.646 & 0.954 & 0.495 & 0.091 & 0.771 & 0.000 & 1.000  \\  
Psychology & 0.812 & 53.220 & 0.000 & 0.186 & 0.920 & -0.460 & 0.562 & 0.129 & 0.904 & -0.112 & 0.475 & -0.791 & 0.208  \\  
Social Sciences & 0.007 & 173.3 & 0.000 & -3.56 & 0.366 & -4.27 & 0.114 & 0.0774 & 0.983 & -1.43 & 0.007 & -4.21 & 0.004  \\  
Veterinary & 0.017 & 38.090 & 0.000 & 0.841 & 0.598 & -0.183 & 0.677 & 1.953 & 0.330 & -0.047 & 0.874 & -0.542 & 0.298  \\  
Dentistry & 0.011 & 11.830 & 0.200 & -6.60 & 0.025 & -6.26 & 0.028 & -3.714 & 0.257 & -1.28 & 0.000 & -7.14 & 0.000  \\  
Health Professions & 0.352 & 38.620 & 0.001 & -0.944 & 0.599 & -1.10 & 0.352 & -0.395 & 0.760 & -0.192 & 0.189 & -1.63 & 0.071  \\  
Multidisciplinary & 0.100 & 98.560 & 0.001 & -1.37 & 0.595 & -1.67 & 0.339 & -1.497 & 0.377 & -0.067 & 0.069 & -1.44 & 0.090  \\  
All Sciences  & 0.076 & 672.3 & 0.000 & 18.30 & 0.009 & -0.125 & 0.797 & -0.007 & 0.992 & -0.054 & 0.625 & -0.240 & 0.488  \\  \hline
\end{tabular}
\begin{flushleft}
Note: Second column gives the \textit{p}-value for the hypothesis that the data follow a power-law model. ``-'' means that the maximum likelihood estimator did not converge. Positive values of the log-likelihood ratio (LR) or the normalized log-likelihood ratio (NLR) indicate that the power-law model is favored over the alternative.
\end{flushleft}
\label{tab:table4}
\end{table}
\end{landscape}

\end{document}